\documentclass[]{article}
\usepackage{bm, amssymb, makecell, graphics, graphicx, mathtools, float, enumitem, booktabs}

\newcommand{\ndkv}{$n$-dimensional $k$-vector}

\newcommand{\T}{^{\mbox{\tiny T}}}

\newcommand{\B}[1]{{\bm #1}}

\usepackage[square,numbers,comma,sort&compress]{natbib}

\begin{document}

\title{Non-dimensional Star-Identification}

\author{Carl Leake\thanks{Aerospace Engineering, Texas A\&M University, College Station, TX 77843-3141, USA. \mbox{leakec@tamu.edu}}, David Arnas\thanks{Postdoctoral Associate, Department of Aeronautics and Astronautics, Massachusetts Institute of Technology, Cambridge, MA. \mbox{arnas@mit.edu}}, and Daniele Mortari\thanks{Aerospace Engineering, Texas A\&M University, College Station, TX 77843-3141, USA. \mbox{mortari@tamu.edu}}}

\date{}
\maketitle

\begin{abstract}This study introduces a new ``Non-Dimensional'' star identification algorithm to reliably identify the stars observed by a wide field-of-view star tracker when the focal length and optical axis offset values are known with poor accuracy. This algorithm is particularly suited to complement nominal lost-in-space algorithms, which may identify stars incorrectly when the focal length and/or optical axis offset deviate from their nominal operational ranges. These deviations may be caused, for example, by launch vibrations or thermal variations in orbit. The algorithm performance is compared in terms of accuracy, speed, and robustness to the Pyramid algorithm. These comparisons highlight the clear advantages that a combined approach of these methodologies provides.
\end{abstract}

\section{Introduction}
If a star tracker is working as intended, a~nominal lost-in-space star identification (Star-ID) algorithm can be used to identify stars using only the observed directions of the unknown stars and the on-board star catalog. These algorithms are paramount for determining the attitude of a spacecraft using a star tracker. In~terms of speed and robustness, the~state-of-the-art algorithm to identify stars in the nominal lost-in-space scenario is the nominal lost-in-space algorithm Pyramid~\cite{Pyramid}. The~Pyramid algorithm recognizes observed stars using the invariance of the angles between observed and cataloged stars. The~$k$-vector range searching technique~\cite{original, Neta} is the internal engine of Pyramid that facilitates a quick and robust Star-ID. The~Pyramid algorithm is summarized in Appendix \ref{app:PyramidSummary}.

While nominal lost-in-space algorithms are typically robust to centroiding error and false stars~\mbox{\cite{Pyramid,Padgett,Luo,Samaan,Liebe,Schiattarella}}, they require an accurate estimate of the star tracker's focal length and optical axis (OA) offset. However, during~a spacecraft's lifetime, environmental effects such as temperature and vibrations may perturb the focal length and OA offset of the star tracker. When this happens, the~Pyramid algorithm may become unable to identify stars or, potentially worse, identify stars incorrectly. Hence the need for a Star-ID algorithm that is less sensitive to these changes in camera~parameters.

To that end, previous algorithms~\cite{ND1,ND2,ND3,NDWideFov,BlindStarID} have proposed solutions based on the non-dimensional Star-ID problem, i.e.,~the identification of stars when the focal length or OA offset of the star tracker were perturbed from their nominal values. However, these algorithms present an important limitation: the non-dimensional database used to obtain the Star-ID must be small. The~small non-dimensional databases are obtained by either filtering the data or by using a star tracker with a narrow field-of-view (FOV). Reference~\cite{ND3} is an example of filtering the data to obtain a small non-dimensional database. The~star catalogue has 1638 stars, but~the star triangles formed by these stars are filtered such that the non-dimensional database contains only 1160 star triangles. Reference~\cite{ND1} is an example of using a small FOV star tracker, $8^{\circ}\times8^{\circ}$, to~obtain a small non-dimensional database. Reference~\cite{ND1} uses the maximum and minimum angles of planar triangles to identify stars. Therefore, if~the FOV of the star tracker becomes too wide, the~resultant non-dimensional star catalog becomes so populated that the difference in angles between adjacent entries are smaller than the centroiding accuracy of the camera. As~a result, it becomes nearly impossible to identify a unique star triangle, and~thus, the~algorithm cannot identify any~stars.

This article proposes the non-dimensional Star-ID algorithm (NDSIA), a~non-dimensional algorithm that is able to handle much larger star catalogs efficiently. This is accomplished by using the dihedral angles of spherical star triangles rather than the angles of planar triangles, and~by considering a database composed of all three spherical triangle angles rather than just one or two of the planar triangles' angles. The term planar triangles refers to triangles that lie on a flat plane. These triangles have familiar properties, such as the sum of their angles is $180^{\circ}$ and the sum of any two sides of the triangle is greater than the third. For~readers unfamiliar with spherical geometry, a~diagram depicting a spherical triangle is shown in Figure~\ref{fig:SphTri} of Appendix \ref{Sec:AppI}. These two changes prevent the non-dimensional catalog from becoming too large; the problem that plagued previous works. As~a consequence of these modifications, the~NDSIA must perform orthogonal range searches in a three-dimensional database, as~opposed to previous algorithms that only dealt with one- or two-dimensional databases. In~general, searching in multidimensional databases is significantly more time-consuming due to the increase in the amount of data, and~the lack of a clear ordering methodology for multidimensional elements (i.e., in~one dimension it is clear how to sort numerical elements so they can be extracted easily, for~example in ascending order, but~in $n$-dimensions such a sorting process is more convoluted). Fortunately, recent advances in the $n$-dimensional $k$-vector~\cite{NDKV} make the searching process viable for real-time applications. As~an example of that, this work shows that the NDSIA takes milliseconds or tens of milliseconds to run, depending on the perturbations of the star tracker~parameters.

The remainder of the article is organized as follows. First, the~theory for the NDSIA is introduced. Then, a~comparison is made between the NDSIA and Pyramid for eight different scenarios, each with different camera perturbations. These eight scenarios showcase the performance of the NDSIA when subject to focal length and OA offset~perturbations.

\section{The Non-Dimensional Star-ID~Algorithm}

The NDSIA is designed for scenarios where the focal length, $f$, and/or OA offset, $[x_{oa}, y_{oa}]$, may~differ from their nominal values, and~is meant to be a backup Star-ID algorithm that is only used when the Pyramid algorithm begins to fail. The~NDSIA is not meant to replace Pyramid when the star tracker is working nominally, as~it requires more memory and a larger computational effort than Pyramid, and~it does not provide any benefits when compared with Pyramid under nominal star tracker conditions. As~an added benefit, once enough stars are identified using the NDSIA, the~actual focal length and OA offset of the camera can be calculated, as~shown in Reference~\cite{ND1}. Once~these quantities are obtained, a~new star database can be generated so that Pyramid can be used again. A~skeleton for implementing Pyramid and the NDSIA synergously is given here, but~the finer details are left to future work, as~the purpose of this article is to introduce and analyze the NDSIA. Typically, the~Star-ID is done by Pyramid, and~the TASTE test~\cite{TASTE}, or~similar self-consistency check, is used to self-diagnose unsuccessful Pyramid Star-IDs (i.e., incorrect star matches). Based on user specifications, if~Pyramid's Star-IDs are unsuccessful often enough, then the Star-ID is performed by the NDSIA instead. The~NDSIA performs the Star-ID until a sufficient number stars have been identified to recalculate the focal length and OA offset of the camera~\cite{ND1}. Finally, a~new database is created based on these parameters, and~the Star-ID is once again performed by~Pyramid.

The non-dimensional algorithm presented here is reminiscent of the algorithms presented in References~\cite{ND1,ND2,ND3,NDWideFov,BlindStarID,Lang}. Nevertheless, the~non-dimensional algorithm introduced in this work has three major differences when compared with these previous~algorithms:
\begin{enumerate}
\item The NDSIA identifies stars using the dihedral angles of spherical star triangles. Previous algorithms used angles of planar star triangles~\cite{ND1,ND2,ND3,NDWideFov}, the~sides of planar star triangles~\cite{BlindStarID}, or~asterisms composed of four or five stars~\cite{Lang}.
\item The NDSIA uses the \ndkv\ (NDKV) \cite{NDKV} to search among a three-dimensional database. Previous algorithms search among a one- or two-dimensional database~\cite{ND1,ND2,ND3,NDWideFov,BlindStarID}, using searching techniques such as the one-dimensional $k$-vector~\cite{Neta}, or~by geometrically hashing asterisms of stars and using a hash-based search algorithm~\cite{Lang}.
\item The NDSIA performs a final check that uses the identified stars' interstellar angles to further the confidence in the Star-ID. This check is non-existent in previous algorithms.
\end{enumerate}

In addition, the~database size (5 GB) and CPU time required by the algorithm of Reference~\cite{Lang} are not well suited for star tracker applications. Conversely, the~database size and CPU time required by NDSIA and algorithms of References~\cite{ND1,ND2,ND3,NDWideFov,BlindStarID} are.

In previous works, planar star triangles were used to identify stars, because~planar triangles are more insensitive to camera perturbations than the interstellar angles used in Pyramid. The~dihedral angles of spherical triangles also have this property. However, the~dihedral angles of spherical triangles contain more information than planar triangle angles, because~planar triangles are the projections on a plane of spherical triangles. A~more mathematically rigorous way of stating this is planar triangle angles are subject to the constraint that the sum of the angles must be equal to $180^{\circ}$. Thus, each~planar triangle only contains two independent pieces of information. In~contrast, spherical triangles do not have this constraint. In~fact, the~sum of spherical triangle angles is in the range $(180^{\circ}, 540^{\circ})$. A~numerical study was performed to identify which positions in an image with a given size and focal length generate the largest sum of the three dihedral angles when projected onto the unit sphere. The~numerical study concluded that the maximum sum of the three dihedral angles occurs when the three stars are on the border of the image, but~their exact positions depend on the image size and the focal length. In~summation, each of the three spherical triangle angles provides an independent piece of information. Hence, more information is available when observing three stars as a spherical triangle rather than as a planar~triangle.

Nevertheless, previous works~\cite{ND1,ND2,ND3,NDWideFov,BlindStarID} were only able to search a one- or two-dimensional database quickly enough for real-time applications. Thus, they were not able to use all the information available from spherical triangles. However, recent advances in the NDKV lifted the one- or two-dimensional database restriction. Consequently, the~information from all three dihedral angles of spherical triangles can be used to uniquely identify stars. Leveraging this information allows for more entries in the non-dimensional database than in previous~works.

\subsection{Creating the Non-Dimensional~Database}

This section presents the database structure used by the NDSIA and shows how it is generated. the NDSIA uses a database, called the non-dimensional database, composed of dihedral angles of spherical star triangles and their corresponding stars. A~visual representation of the dihedral angles of spherical star triangles and a detailed description of how they are calculated is shown in Appendix \ref{Sec:AppI}. To~construct the non-dimensional database, the~algorithm needs to first identify all the potential star triangles that fit within the FOV of the star tracker. Once these admissible triangles have been identified, their dihedral angles are computed using the equations shown in Appendix \ref{Sec:AppI}. Each spherical triangle becomes an element in the non-dimensional database, where each element is composed of the dihedral angles, sorted in ascending order, followed by the stars that lie at the vertices associated with the sorted dihedral~angles.

As mentioned previously, the~NDSIA requires more memory than Pyramid. A~comparison of the databases used by each of theses algorithms quantifies this statement. Both databases were created in Python on Ubuntu 18.04 with an Intel(R) Core(TM) i5-2400 CPU at 3.10 GHz and 16.0 GB of RAM. Each~database was generated from an original star catalog of 1673 stars. The~non-dimensional database took 901 seconds to generate, and~contains 6 rows and 2,762,895 columns. It takes up 99.46 MB of memory. The~Pyramid database took 0.175 seconds to generate, and~contains 99,422 rows and 3~columns. It takes up 1.59 MB of memory, almost two orders of magnitude less memory than the NDSIA. While the database generation can be done offline, so the computational time is not necessarily important, it is included here for~completeness. 

\subsection{The Non-Dimensional Star-ID~Algorithm}
\begin{figure}[!ht]
    \centering
    \includegraphics[width=0.90\linewidth]{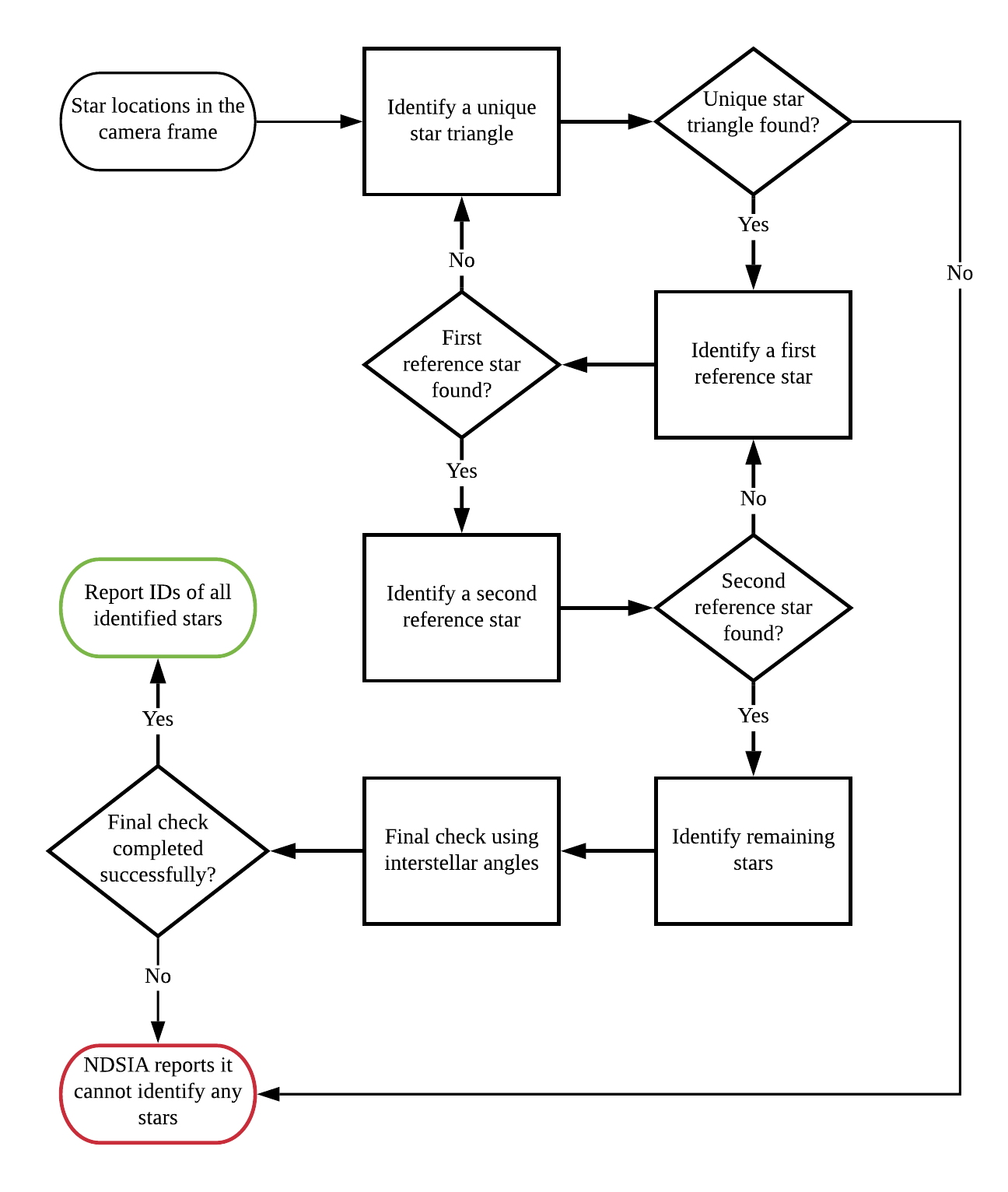}
    \caption{The NDSIA~Flowchart.}
    \label{fig:NdsiaFlowchart}
\end{figure}

The identification process of the NDSIA is similar to that of Pyramid~\cite{Pyramid}, but~with two key differences. First, the~NDSIA makes use of two reference stars rather than one, and~second, the~NDSIA uses a final check at the end of the identification process that Pyramid does not. The~flowchart in Figure~\ref{fig:NdsiaFlowchart} summarizes the NDSIA algorithm. The~rectangular boxes in Figure~\ref{fig:NdsiaFlowchart} represent processes. Each process contains a set of specific steps that are described in detail in the subsections that~follow.
\unskip

\subsubsection{Identify a Unique Star~Triangle}

A unique star triangle is a star triangle that only returns one match when searched for in the non-dimensional star database. A~star triangle is identified by its dihedral angles. Thus, to~find a unique star triangle, the~dihedral angles must be computed using the equations in Appendix \ref{Sec:AppI}. The~dihedral angles are sorted in ascending order, to~match the non-dimensional database, and~an orthogonal range is created by adding and subtracting the quantity $\epsilon$ from the dihedral angles. Then, the~NDKV orthogonal range searching algorithm~\cite{NDKV} is used to find all possible triangles in the database that fit this range. The~parameter $\epsilon$ can be used to tune the performance of the NDSIA. If~$\epsilon$ is smaller, the~Star-ID process will be completed a lower percentage of the time, but~the confidence in the Star-ID increases. If~the parameter $\epsilon$ is larger, the~opposite is true. In~this work, $\epsilon$ is chosen to be three standard deviations of the star tracker's centroiding error. If~the range search only returns one match and the $L_2$ norm of the difference between the spherical triangle angles in the camera frame and those in the database is less than $\epsilon$, then that triangle is a unique star triangle. The~latter of these conditions can be mathematically expressed as,
\begin{equation}\label{eq:DiHedAngTol}
    \sqrt{(A-A_d)^2+(B-B_d)^2+(C-C_d)^2} < \epsilon,
\end{equation}
where $A$, $B$, and~$C$ are the dihedral angles computed in the camera frame and $A_d$, $B_d$, and~$C_d$ are the angles of the matching spherical triangle in the non-dimensional database. If~the range search returns zero matches, more than one match, or~the inequality in Equation~\eqref{eq:DiHedAngTol} is not satisfied, then that triangle is not a unique star~triangle.

All possible combinations of three stars in the star tracker's FOV are tested until a unique star triangle is identified. The~method used to test all three star combinations is known as the star kernel generator. Naturally, one may be inclined to use a star kernel generator that loops through the stars in order to identify a unique star triangle. However, this is not the most efficient method. For~example, if~the first star in the loop is a false-star (i.e., noise in the camera rather than a true star), then all triangles using that star will have to be tested before moving onto the next star. Hence, a~more efficient star kernel generator is employed to ensure that all triangles are tested, but~that avoids using one particular star over and over again: the pattern shifting star kernel generator from Reference~\cite{triads}. Please note that the enhanced pattern shifting star kernel generator from Reference~\cite{triads} was tested as well, but~did not show any noticeable speed improvements over the pattern shifting star kernel generator for the numerical tests of the NDSIA shown in this~article. 

If no unique star triangle is found, then the NDSIA exits and reports it cannot identify any stars. If~a reference star is found, then the algorithm moves onto the next step. Let the candidate stars of the unique star triangle found in this step be denoted as the $i$, $j$, and~$k$ stars. Let the unique star triangle itself be referred to as the $\{i,j,k\}$ triangle.

\subsubsection{Identify a First Reference~Star}\label{sec:firstref}

To minimize the chance of identifying stars incorrectly, two reference stars are used to verify the $i$, $j$, and~$k$ stars. Let $r$ denote the first reference star. The~first reference star is considered identified if and only if the star triangles $\{r,i,j\}$, $\{r,i,k\}$, and~$\{r,j,k\}$ are unique star triangles. All stars that are not the $i$, $j$, and~$k$ stars are tested as potential reference stars until one is found. If~a reference star is found, then the algorithm continues on to the next step. Otherwise, the~algorithm moves back a step and continues trying to find a different set of three stars to use as the $\{i,j,k\}$ triangle.

\subsubsection{Identify a Second Reference~Star}

To minimize the chance of identifying stars incorrectly, a~second reference star is used to verify the $i$, $j$, $k$, and~$r$ stars. Let $r_2$ denote the second reference star. The~second reference star is considered identified if and only if the star triangles $\{i,j,r_2\}$, $\{i,k,r_2\}$, $\{j,k,r_2\}$, $\{i,r,r_2\}$, $\{j,r,r_2\}$, and~$\{k,r,r_2\}$, are unique star triangles. All stars that are not the $i$, $j$, $k$, and~$r$ stars are tested as potential second reference stars until one is found. If~a second reference star is found, then all five stars, $i$, $j$, $k$, $r$, and~$r_2$, are considered identified and are recorded. Otherwise, the~algorithm moves back a step and continues trying to find a different first reference star $r$.

\subsubsection{Identify Remaining~Stars}

After identifying the $i$, $j$, $k$, $r$, and~$r_2$ stars, the~remaining stars in the frame are identified using the same process as the first reference star. That is, any remaining star $s$ can be identified if and only if the star triangles $\{s,i,j\}$, $\{s,i,k\}$, and~$\{s,j,k\}$ are unique star triangles. If~the aforementioned triangles are not unique star triangles, then star $s$ is~discarded. 

\subsubsection{Final Check Using Interstellar~Angles}

Once all the stars that can be identified have been identified, a~final check is performed using the identified stars' interstellar angles. This is done by calculating the interstellar angles between every possible star pair from the subset of stars already identified. Appendix \ref{Sec:AppI} shows how to calculate the interstellar angle between two stars. If~any of the interstellar angles obtained is greater than the FOV of the camera, then an error has been made in the Star-ID process. If~this happens, then no stars can be accurately identified, and~thus, the~algorithm reports that it cannot identify any~stars.

\section{The Non-Dimensional Star-ID Algorithm Compared with~Pyramid}

In this section, the~NDSIA is compared with Pyramid using the following five~metrics:
\begin{enumerate}
\item Number of scenes wherein the Star-ID was completed. This metric shows whether the algorithm attempted to perform a Star-ID or not. This metric is reported as a percentage of the total number of scenes in the test.
\item Number of successful Star-IDs. This metric is calculated as the ratio of the number of times that all stars identified by the algorithm were identified correctly, to~the number of times the Star-ID was completed. This ratio is reported as a percentage.
\item Average time that it takes to perform one Star-ID. This metric is reported in milliseconds.
\item Number of scenes wherein Pyramid did not complete the Star-ID and the NDSIA completed the Star-ID successfully. This metric is reported as a percentage of the number scenes wherein Pyramid did not complete the Star-ID.
\item Number of scenes wherein Pyramid completed the Star-ID unsuccessfully (i.e., at~least one star was identified incorrectly) and the NDSIA completed the Star-ID successfully. This~metric is reported as a percentage of the number of scenes wherein Pyramid completed the Star-ID~unsuccessfully.
\end{enumerate}

These five metrics are used to evaluate the NDSIA's performance over a set of eight tests designed to cover different star tracker situations. Table~\ref{tab:testDescriptions} describes the parameters used for each~test.
\begin{table}[H]
    \centering
    \caption{Test~descriptions.}\label{tab:testDescriptions}
    \scalebox{.85}[0.85]{\begin{tabular}{cccc}
        \toprule
        \textbf{Test Number} & \makecell{\textbf{Focal Length}\\\textbf{Perturbation}\\\textbf{(\% of Nominal Focal Length)}} & \makecell{\textbf{Optical Axis}\\\textbf{Offset Perturbation}\\\textbf{(\% of Half the Imager Width)}} & \makecell{\textbf{Camera Centroiding}\\\textbf{Error Standard}\\\textbf{Deviation (Arcsec)}} \\
        \midrule
        1 & 0 & 0 & 10 \\
        \midrule
        2 & 0.5 & 0 & 10 \\
        \midrule
        3 & 2.0 & 0 & 10 \\
        \midrule
        4 & 0 & 0.5 & 10 \\
        \midrule
        5 & 0 & 2.0 & 10 \\
        \midrule
        6 & 0.5 & 0.5 & 10 \\
        \midrule
        7 & 2.0 & 2.0 & 10 \\
        \midrule
        8 & 0.5 & 0.5 & 15 \\
        \bottomrule
    \end{tabular}}
    
\end{table}
\unskip

\subsection{Test~Conditions}

Each test contained 1000 randomly oriented scenes, where a scene consists of performing the Star-ID and estimating the attitude. The~random orientations were generated via $QR$ decomposition of a randomly generated $3\times3$ matrix. The~seed used to generate the random $3\times3$ matrix was fixed such that all tests used the same random orientations. All of the tests included in this section were performed in C++ on a computer running Ubuntu 18.04 with an Intel(R) Core(TM) i5-2400 CPU at 3.10 GHz and 16.0 GB of RAM. All run times were calculated using the system\_clock function in the C++ boost chrono library. The~parameters of the virtual star tracker used in these tests are shown in Table~\ref{tab:StarTrackerParams}, where $U[0,5]$ represents the uniform distribution of integers in the range $[0,5]$ and $\sigma$ denotes the standard deviation of a normal~distribution.

\begin{table}[H]
    \centering
    \caption{Star tracker~parameters.}\label{tab:StarTrackerParams}
    \begin{tabular}{ll}
        \toprule
        \multicolumn{1}{c}{\textbf{Virtual Star Tracker Parameter}} & \multicolumn{1}{c}{\textbf{Value}}\\ \midrule
        $1\sigma$ centroid error & $10$ arcseconds \\ \midrule
        Star magnitude threshold & $5.0$ \\ \midrule
        Number of false stars & $U[0,5]$ \\ \midrule
        Number of pixel rows & $1024$ \\ \midrule
        Number of pixel columns & $1024$ \\ \midrule
        Pixel pitch & $0.018$ mm \\ \midrule
        Focal length & $50.47$ mm \\ \midrule
        Number of stars in star catalogue & 1673\\ \midrule
        Number of interstellar angles in Pyramid database & 99,422 \\ \midrule
        Number of star triangles in non-dimensional database & 2,762,895 \\ \bottomrule
    \end{tabular}
\end{table}

The centroiding error was applied to the star measurements in the camera frame using the methodology shown in Equation~\eqref{eq:CentroidError}.
\begin{align}\label{eq:CentroidError}
    \theta &\sim \mathcal{N}(0,\sigma^2)\nonumber\\
    \B{e} &= \B{v}\times \hat{\B{b}}_t\nonumber\\
    \hat{\B{e}} &= \B{e}/||\B{e}||\\
    C_p &= C_p(\hat{\B{e}},\theta)\nonumber\\
    \hat{\B{b}}_e &= C_p \, \hat{\B{b}}_t\nonumber
\end{align}
where $\mathcal{N}(\mu,\sigma^2)$ is the normal distribution with mean $\mu$ and variance $\sigma^2$, $\sigma$ is the centroid error of the camera, $\hat{\B{b}}_t\in\mathcal{R}^3$ is a unit vector that points in the true direction of the star in the camera reference frame, $\hat{\B{b}}_e\in\mathcal{R}^3$ is the observed unit vector, affected by centroid error, $\B{v}\in\mathcal{R}^3$ is a random vector, and~the function $C_p (\hat{\B{e}}, \theta)$ produces an attitude matrix with principle axis $\hat{\B{e}}$ and principle angle $\theta$. The~function used to calculate $C_p$ given $\hat{\B{e}}$ and $\theta$ is,
\begin{equation*}
    C_p (\hat{\B{e}}, \theta) = \mathcal{I}_{3\times 3} \, \cos\theta + (1 - \cos\theta) \, \hat{\B{e}} \hat{\B{e}}\T - [\hat{\B{e}}\times] \, \sin\theta
\end{equation*}
where $\mathcal{I}_{3\times 3}$ is the $3\times 3$ identity matrix and $[\hat{\B{e}}\times]$ is the skew-symmetric matrix formed using the components of $\hat{\B{e}}$.

The focal length and OA offset perturbations are applied to the star measurements in the camera frame using the methodology shown in Equation~\eqref{eq:CameraPerturbations},
\vspace{-3pt}
\begin{align}\label{eq:CameraPerturbations}
    \begin{Bmatrix} x_c \\ y_c \end{Bmatrix} &= \begin{bmatrix} -f-\delta f & 0 & 0 \\ 0 & -f -\delta f & 0\end{bmatrix}\frac{\hat{\B{b}}_e}{\hat{\B{b}}_e(3)}\nonumber \\
    \B{b}_p &= \begin{Bmatrix} \pm x_c - x_{oa} + \delta x \\ \pm y_c - y_{oa}+\delta x \\ f + \delta f\end{Bmatrix}\\
    \hat{\B{b}}_p &= \dfrac{\B{b}_p}{| \, \B{b}_p \, |}\nonumber,
\end{align}
where $\hat{\B{b}}_e(3)$ is the third component of the vector $\hat{\B{b}}_e$, $\delta f$ is the focal length perturbation, $\delta x$ and $\delta y$ are the OA offset perturbations, and~$\hat{\B{b}}_p$ is the unit vector pointing to the star whose centroid coordinates are $[x_c, y_c]$ on the imager. The~``$\pm$'' sign appearing in Equation~\eqref{eq:CameraPerturbations} depends on the imager $x$ and $y$ axes directions. Please note that Equation~\eqref{eq:CameraPerturbations} assumes that the star tracker is modelled as an ideal pin-hole~camera. 

\subsection{Results}

Table~\ref{tab:ndTests} gives a summary of the performance and results of Pyramid and the NDSIA on the eight tests. For~each algorithm, three columns are included that contain the first three metrics described earlier. The~first column, $n_{id} (\%)$, gives the percentage of scenes where the Star-ID was completed, and~the second column, $n_{+id}$, shows the percentage of completed Star-ID runs with a successful Star-ID (i.e., a~Star-ID wherein all stars identified by the algorithm are identified correctly). The~third column shows the average time each algorithm took to perform the Star-ID in milliseconds. In~the joint statistics column of the table, the~remaining two metrics are~reported.

\begin{table}[H]
\caption{Comparison between the Pyramid algorithm and the~NDSIA.}
\begin{center}
\begin{tabular}{ccccccccc}
\toprule
\makecell{\textbf{Test}} & \multicolumn{3}{c}{\textbf{Pyramid}} & \multicolumn{3}{c}{\textbf{NDSIA}} & \multicolumn{2}{c}{\textbf{Joint Statistics}} \\ \midrule
 & \makecell{\boldmath{$n_{id}$}\\\textbf{(\%)}} & \makecell{\boldmath{$n_{+id}$}\\\textbf{(\%)}} & \makecell{\boldmath{$t_{avg}$}\\\textbf{(ms)}} & \makecell{\boldmath{$n_{id}$}\\\textbf{(\%)}} & \makecell{\boldmath{$n_{+id}$}\\\textbf{(\%)}} & \makecell{\boldmath{$t_{avg}$}\\\textbf{(ms)}} & \makecell{\textbf{Incomplete}\\\textbf{Pyramid Star-ID}\\\textbf{Scenes Completed}\\\textbf{Successfully by}\\\textbf{the NDSIA (\%)}} & \makecell{\textbf{Unsuccessful}\\\textbf{Pyramid Star-ID}\\\textbf{Scenes Completed}\\\textbf{Successfully by}\\\textbf{the NDSIA (\%)}}\\
\midrule
1 & 100 & 99.9 & 0.139 & 84.2 & 100 & 2.81 & N/A 
 & 100 \\
\midrule
2 & 41.8 & 90.7 & 10.8 & 79.4 & 100 & 3.80 &
68.0 & 92.3  \\
\midrule
3 & 9.20 & 41.3 & 27.8 & 21.5 & 100 & 24.3 &
17.8 & 59.3 \\
\midrule
4 & 100 & 99.9 & 0.141 & 81.8 & 100 & 3.40 & N/A
 & 100 \\
\midrule
5 & 100 & 99.9 & 0.140 & 53.5 & 100 & 12.6 & N/A
 & 100 \\
\midrule
6 & 41.7& 89.9 & 10.9 & 77.2 & 100 & 4.32 &
64.5 & 95.2 \\
\midrule
7 & 10.2 & 38.2 & 27.6 & 14.7 & 100 & 29.4 &
10.6 & 54.0 \\
\midrule
8 & 75.5 & 90.3 & 6.92 & 75.9 & 100 & 7.06 &
46.5 & 91.8 \\
\bottomrule
\end{tabular}
\end{center}
\label{tab:ndTests}
\end{table}

In general, Table~\ref{tab:ndTests} shows that the NDSIA is reliable, even when the camera parameters are perturbed, as~it never completes a Star-ID unsuccessfully (i.e., it never incorrectly identifies a star). Furthermore, in~every test, the~NDSIA is able to successfully complete the Star-ID in more than 50\% of the scenes wherein Pyramid completes the Star-ID unsuccessfully. The~joint statistics for the NDSIA completing Star-IDs successfully where Pyramid does not complete the Star-ID at all are less impressive, but~still show that the NDSIA maintains some benefit in this regard over Pyramid in all tests with focal length perturbations. Finally, Table~\ref{tab:ndTests} shows that the NDSIA takes on the order of tens of milliseconds or less to perform the Star-ID, a~speed that is suitable for many real-time~applications.

Test 1 shows that when the star tracker is working nominally Pyramid and the NDSIA are each able to identify the stars correctly, with~one exception; in one case, Pyramid performs the Star-ID incorrectly. However, in~this case, the~final attitude error, $0.011$ degrees, is still small. Pyramid is able to identify stars more often than the NDSIA, and~Pyramid is more than an order of magnitude faster than the NDSIA. This is expected, as~the NDSIA is meant to be used only when the Pyramid algorithm begins to~fail. 

Test 2 and Test 3 are the first two examples of such situations. These tests were performed with focal length perturbations. Compared to the nominal case, the~Pyramid algorithm performance degrades significantly. The~most alarming change in the performance of the Pyramid algorithm is the reduction in the percentage of scenes with a successful Star-ID. In~Test 2 and Test 3, the~Star-ID performed by Pyramid can no longer be trusted. In~contrast, the~percentage of the NDSIA scenes with a successful Star-ID remains unchanged from the nominal case. However, the~number of Star-IDs that can be completed is reduced significantly when compared with the nominal case. In~addition, the~time required to complete a Star-ID increases due to the focal length~perturbations. 

In Test 4 and Test 5, the~performance of the Pyramid algorithm remains almost unchanged when compared with Test 1, the~nominal case. The~reason is small OA offset perturbations have little effect on the interstellar angles between stars for most locations on the imager. Therefore, the~Pyramid algorithm is still able to perform the Star-ID. Conversely, the~small OA offset perturbations have a larger effect on the dihedral angles between the stars. Thus, the~number of times that the Star-ID can be completed by the NDSIA is reduced as the OA offset increases. However, the~percentage of scenes with a successful Star-ID remains unchanged from the nominal case. Therefore, while the NDSIA may not be able to perform the Star-ID process as often as Pyramid when there is only an OA offset perturbation, the~Star-ID provided by the NDSIA can still be~trusted.

Test 6 and Test 7 show the performance of the two algorithms when subject to both focal length perturbations and OA offset perturbations. In~these two tests, there is again a significant degradation in the performance of the Pyramid algorithm, and~the number of Star-IDs that both algorithms can complete is reduced when compared with Test 1, the~nominal~case.

Test 8 shows that the NDSIA is robust to changes in the centroiding accuracy of the camera, because~the performance of the NDSIA in Test 8 is similar to the performance of the NDSIA in Test~6---Test 6 is the same as Test 8 but with a lower centroiding error. Comparing Pyramid and the NDSIA on Test 8 reveals that the NDSIA does better in terms of percentage of Star-IDs completed and percentage of Star-IDs completed successfully. The~Pyramid algorithm is slightly faster than the NDSIA in Test 8. Comparing the Pyramid results for Test 8 with the Pyramid results for Test 6 shows that Pyramid is able to perform more Star-IDs in Test 8 than Test 6. The~larger centroiding error absorbs some of the error due to the focal length and OA offset perturbations. In~other words, in~many scenes, the~position error of stars in the camera frame due to focal length and OA offset perturbations is less than the typical centroiding error in Test 8. Thus, with~the increased centroiding error, and~therefore an increased range when performing the associated range searches, Pyramid is able to identify stars in more~scenes.

Based on the results of Test 8, one is naturally led to wonder if the performance of Pyramid can be improved by increasing the range it uses when performing range searches in its database. Please note that in Test 8 both the actual camera centroiding error and the range used in Pyramid were modified. In~contrast, here, only the range Pyramid uses will be modified, the~actual camera centroiding error will be held constant, using the value given in Table~\ref{tab:StarTrackerParams}. Table~\ref{tab:testPyramidSig} shows how the performance of Pyramid is impacted by changing the size of the range search. The~left most column in Table~\ref{tab:testPyramidSig} gives the standard deviation, $\sigma$, that Pyramid assumes the star tracker camera has; hence, the~range that Pyramid uses when searching the database of interstellar angles is $\pm3\sigma$. The~focal length and OA offset perturbations used to create Table~\ref{tab:testPyramidSig} are the same as those used in Test 6. Hence, the~first row of Table~\ref{tab:testPyramidSig} is identical to that of Test 6. Each row in Table~\ref{tab:testPyramidSig} was created using 1000 scenes, the~same as in the previous eight~tests.
\begin{table}[H]
    \centering
    \caption{Pyramid performance as a function of range search~size.}\label{tab:testPyramidSig}
    \begin{tabular}{cccc}
        \toprule
        \makecell{\boldmath{$\sigma$} \textbf{Used for Pyramid}\\\textbf{Range Search (Arcsec)}}  & \boldmath{$n_{id}$} \textbf{(\%)} & \boldmath{$n_{+id}$} \textbf{(\%)} & \boldmath{$t_{avg}$} \textbf{(ms)}  \\
        \midrule
        10 & 41.7 & 89.9 & 10.9\\
        \midrule
        15 & 74.6 & 90.5 & 7.16\\
        \midrule
        20 & 93.3 & 90.7 & 3.63\\
        \midrule
        25 & 98.2 & 88.8 & 2.65\\
        \midrule
        30 & 99.3 & 90.0 & 2.93\\
        \midrule
        35 & 99.8 & 89.4 & 4.05\\
        \midrule
        40 & 99.9 & 89.8 & 7.26\\
        \bottomrule
    \end{tabular}
\end{table}
Table~\ref{tab:testPyramidSig} shows that as the range that Pyramid uses increases, so does the percentage of tests with completed Star-IDs. Moreover, the~percentage of successful star IDs remains approximately constant. As~a result, the~total number of tests for which Pyramid completes a successful Star-ID increases as the range increases. However, this also means that there are still a notable number of cases for which Pyramid identifies stars incorrectly. Hence, even by varying the range that Pyramid uses to search in the database, the~Pyramid algorithm does not outperform the~NDSIA. 

\begin{figure}[H]
    \centering
    \includegraphics[width=0.7\linewidth]{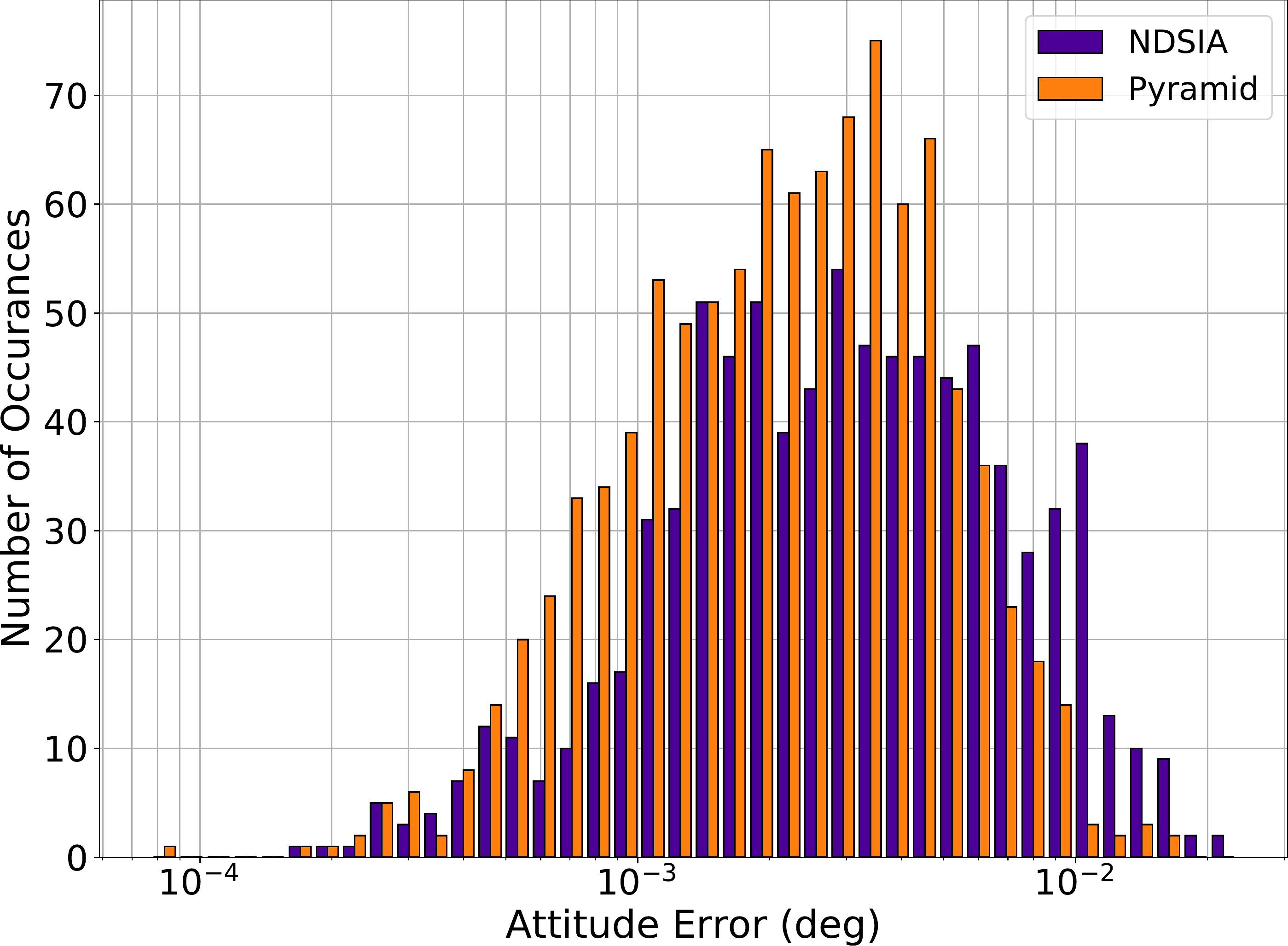}
    \caption{Test 1, Nominal Test, Attitude Estimation Error for Scenes with a Successful~Star-ID.}
    \label{fig:test1}
\end{figure}

Histograms of attitude error are provided for each test to quantify how much the perturbations ultimately affect the attitude estimation. The~histograms only include the attitude estimation error for scenes where the Star-ID was completed successfully, as~almost all scenes with an incorrect Star-ID have a large attitude error. The~q-method is used to estimate the attitude for each algorithm~\cite{qMethod}. Therefore, the~only variable that affects the attitude estimation error is the number of stars identified by each algorithm in each scene. Figure~\ref{fig:test1} shows the histograms of attitude error for Pyramid and the NDSIA for Test 1. This figure shows that the two algorithms have similar attitude error distributions, but~that Pyramid completed more Star-IDs than the~NDSIA.

\begin{figure}[H]
    \centering
    \includegraphics[width=0.75\linewidth]{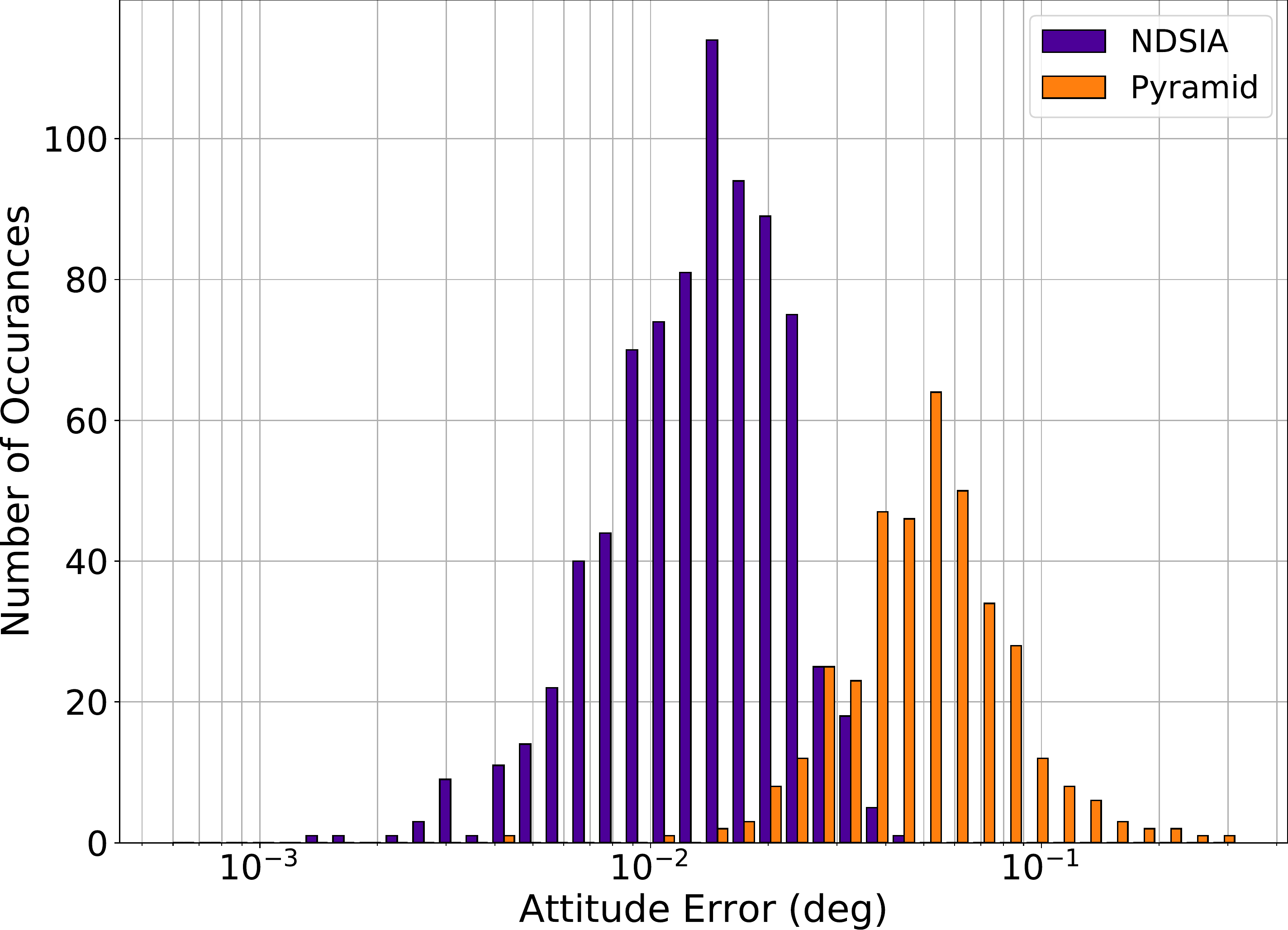}
    \caption{Test 2, Small Focal Length Perturbation, Attitude Estimation Error for Scenes with a Successful~Star-ID.}
    \label{fig:test2}
\end{figure}

\begin{figure}[H]
    \centering
    \includegraphics[width=0.75\linewidth]{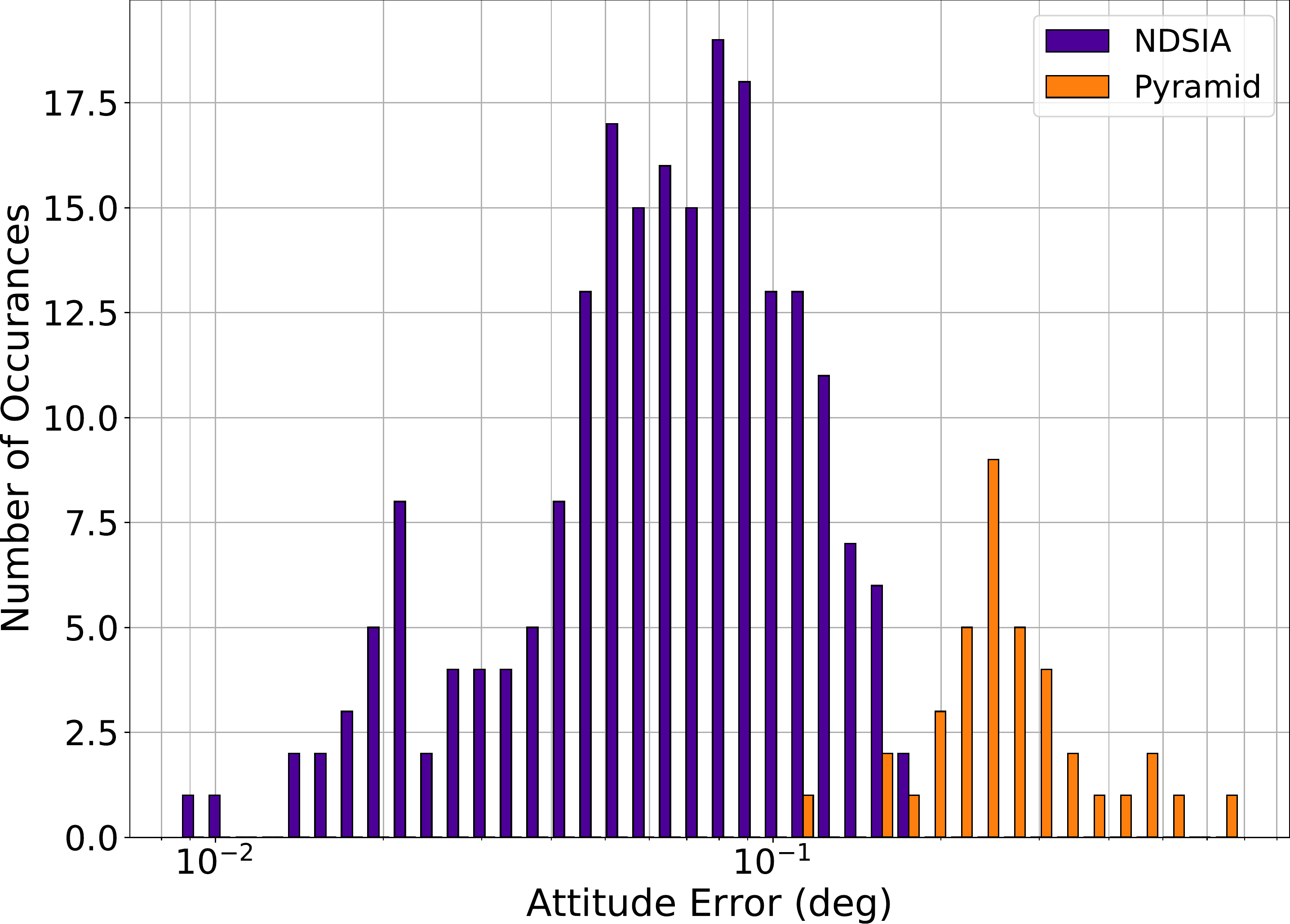}
    \caption{Test 3, Large Focal Length Perturbation, Attitude Estimation Error for Scenes with a Successful~Star-ID.}
    \label{fig:test3}
\end{figure}

Figures~\ref{fig:test2} and \ref{fig:test3} show histograms of the attitude error for Pyramid and the NDSIA for Test 2 and Test 3. These histograms clearly show that the NDSIA is more accurate on average than the Pyramid algorithm. Moreover, comparing Test 2 and Test 3 to Test 1 reveals that the attitude error of the NDSIA increases slightly as the focal length perturbation~increases.
\unskip

\begin{figure}[H]
    \centering
    \includegraphics[width=0.75\linewidth]{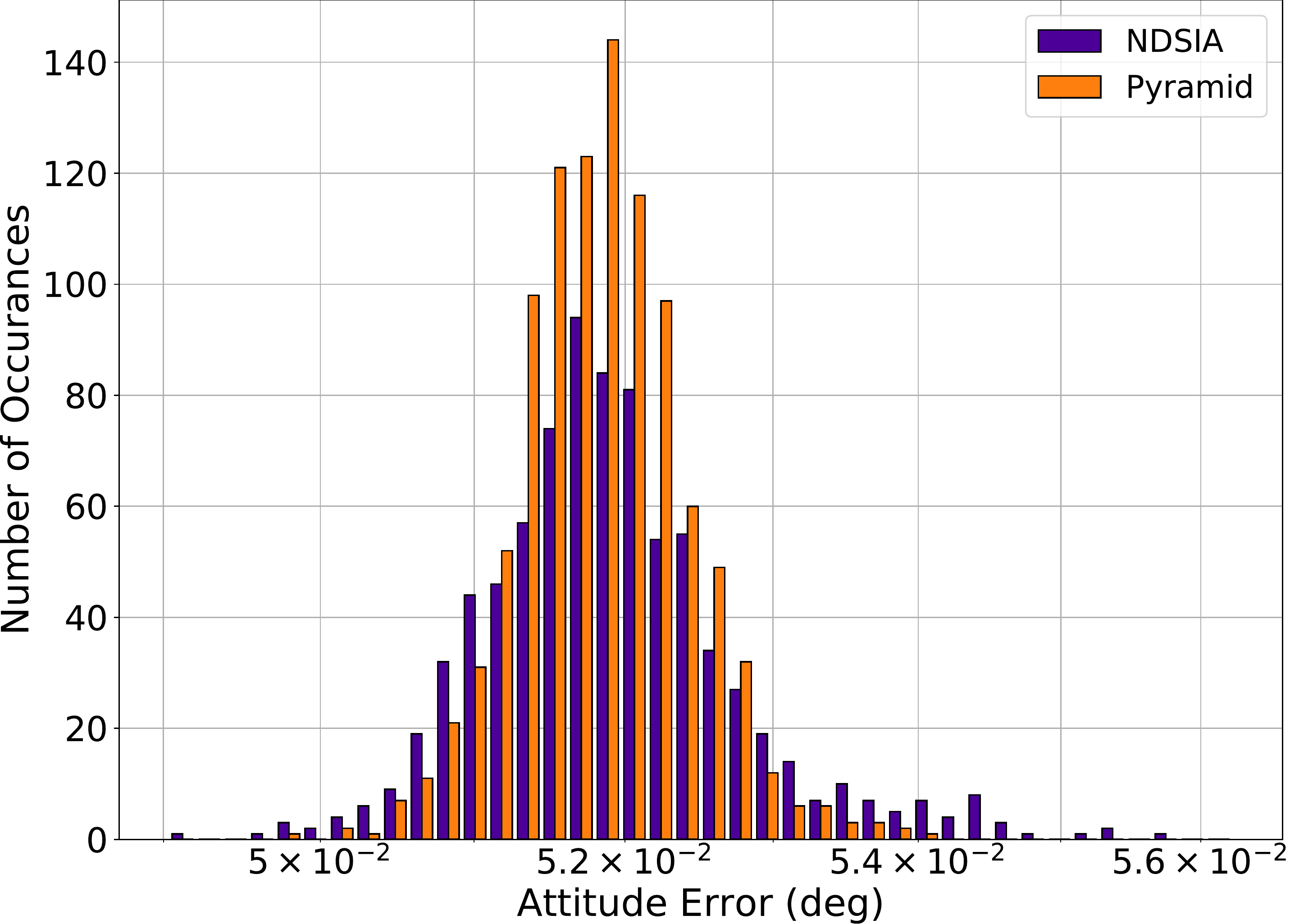}
    \caption{Test 4, Small OA Offset Perturbation, Attitude Estimation Error for Scenes with a Successful~Star-ID.}
    \label{fig:test4}
\end{figure}

\begin{figure}[H]
    \centering
    \includegraphics[width=0.75\linewidth]{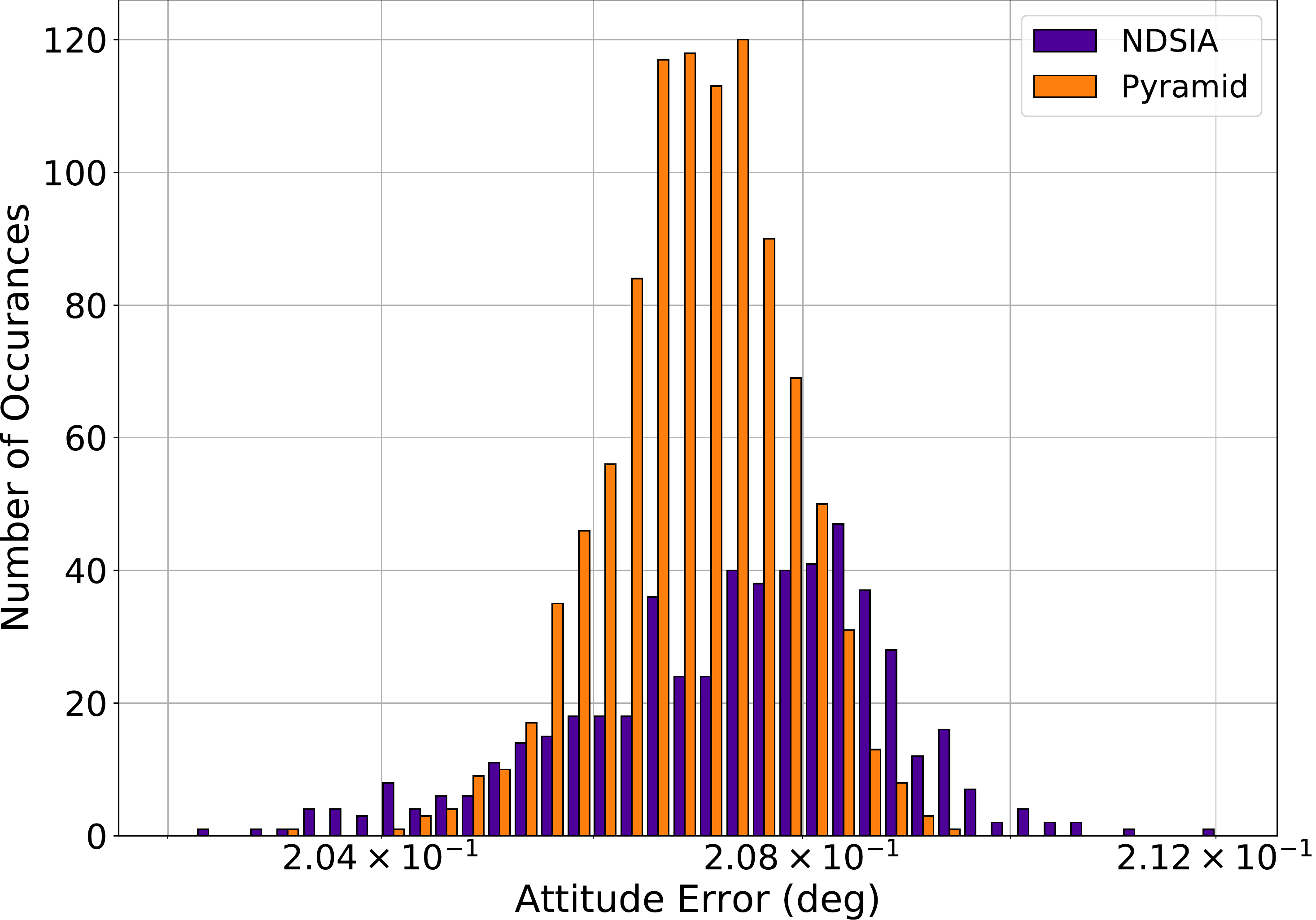}
    \caption{Test 5, Large OA Offset Perturbation, Attitude Estimation Error for Scenes with a Successful~Star-ID.}
    \label{fig:test5}
\end{figure}

Figures~\ref{fig:test4} and \ref{fig:test5} show the attitude error histograms for Pyramid and the NDSIA for Test 4 and Test 5. These figures show that the attitude error distributions for the two algorithms are similar for both tests; however, in~each case, the~NDSIA completed fewer Star-IDs than the Pyramid algorithm. When compared to Test 1, the~attitude error of both algorithms increases with the increase in the OA offset~perturbation.
\unskip

\begin{figure}[H]
    \centering
    \includegraphics[width=0.75\linewidth]{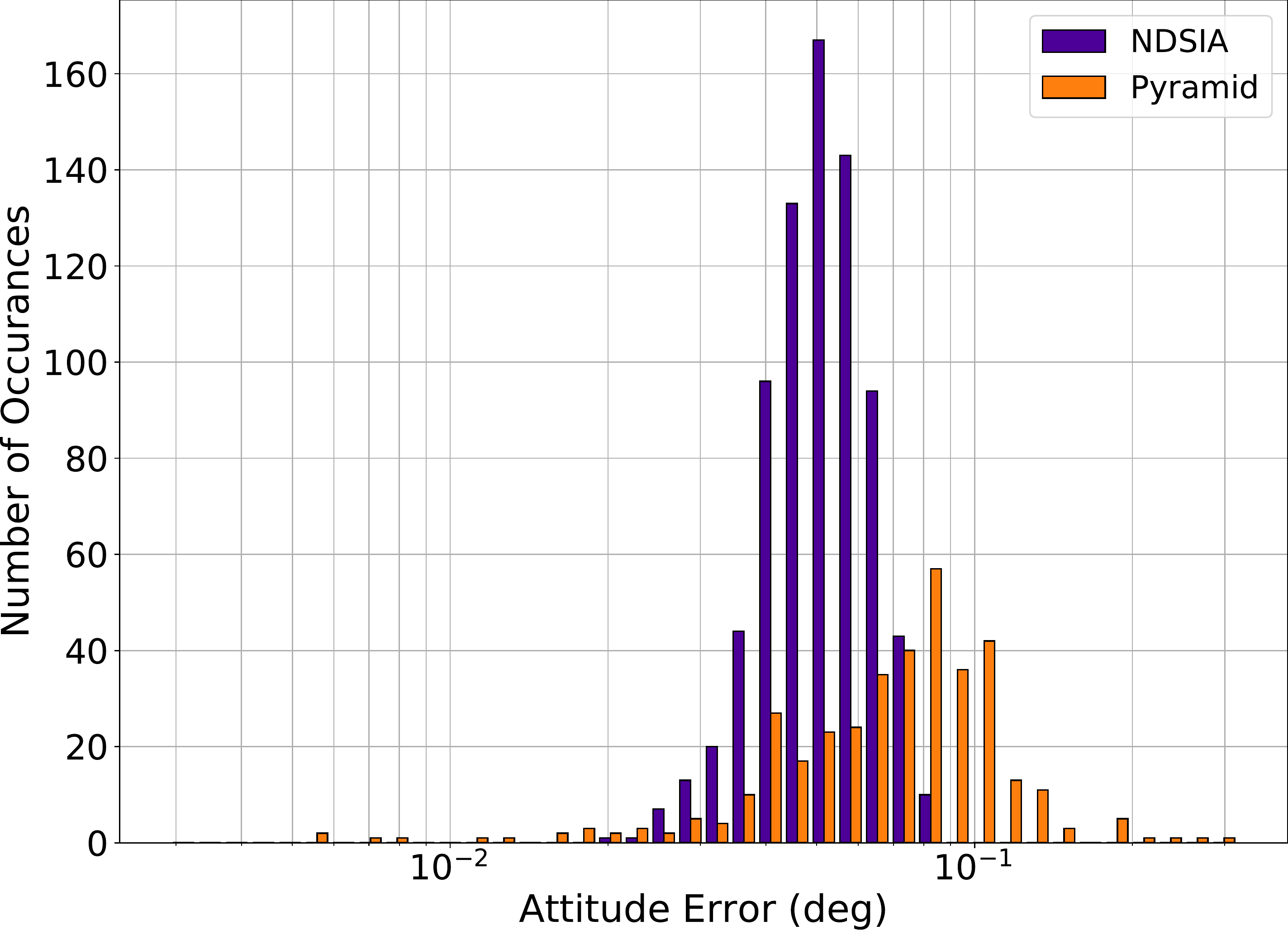}
    \caption{Test 6, Small Focal Length and OA Offset Perturbation, Attitude Estimation Error for Scenes with a Successful~Star-ID.}
    \label{fig:test6}
\end{figure}

\begin{figure}[H]
    \centering
    \includegraphics[width=0.75\linewidth]{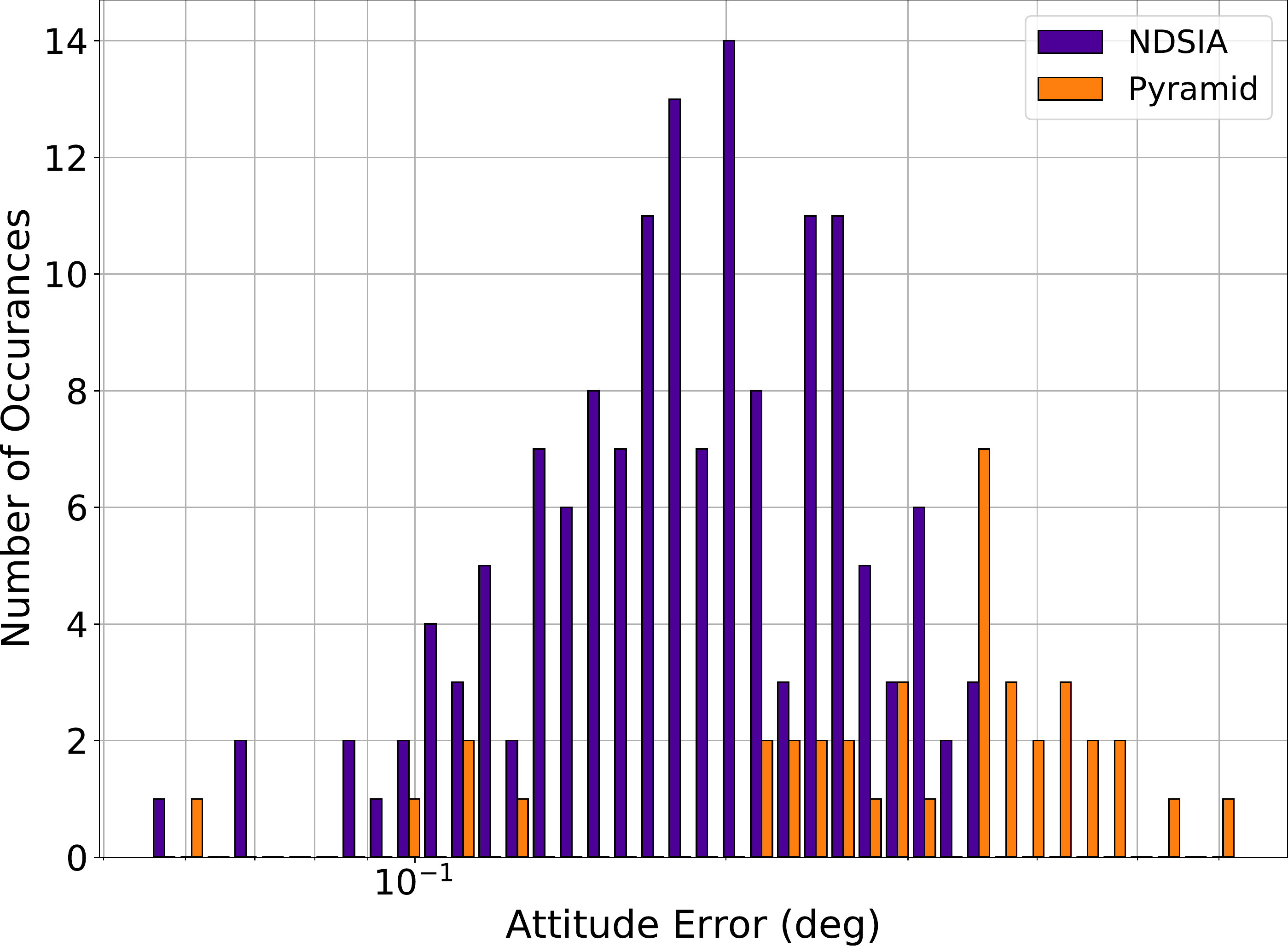}
    \caption{Test 7, Large Focal Length and OA Offset Perturbation, Attitude Estimation Error for Scenes with a Successful~Star-ID.}
    \label{fig:test7}
\end{figure}

Figures~\ref{fig:test6} and \ref{fig:test7} show the attitude error histograms for Pyramid and the NDSIA for Test 6 and Test 7. These figures show that on average the NDSIA is more accurate than Pyramid. In~these tests, the~NDSIA is able to complete the Star-ID process in more cases than~Pyramid.
\unskip

\begin{figure}[H]
    \centering
    \includegraphics[width=0.75\linewidth]{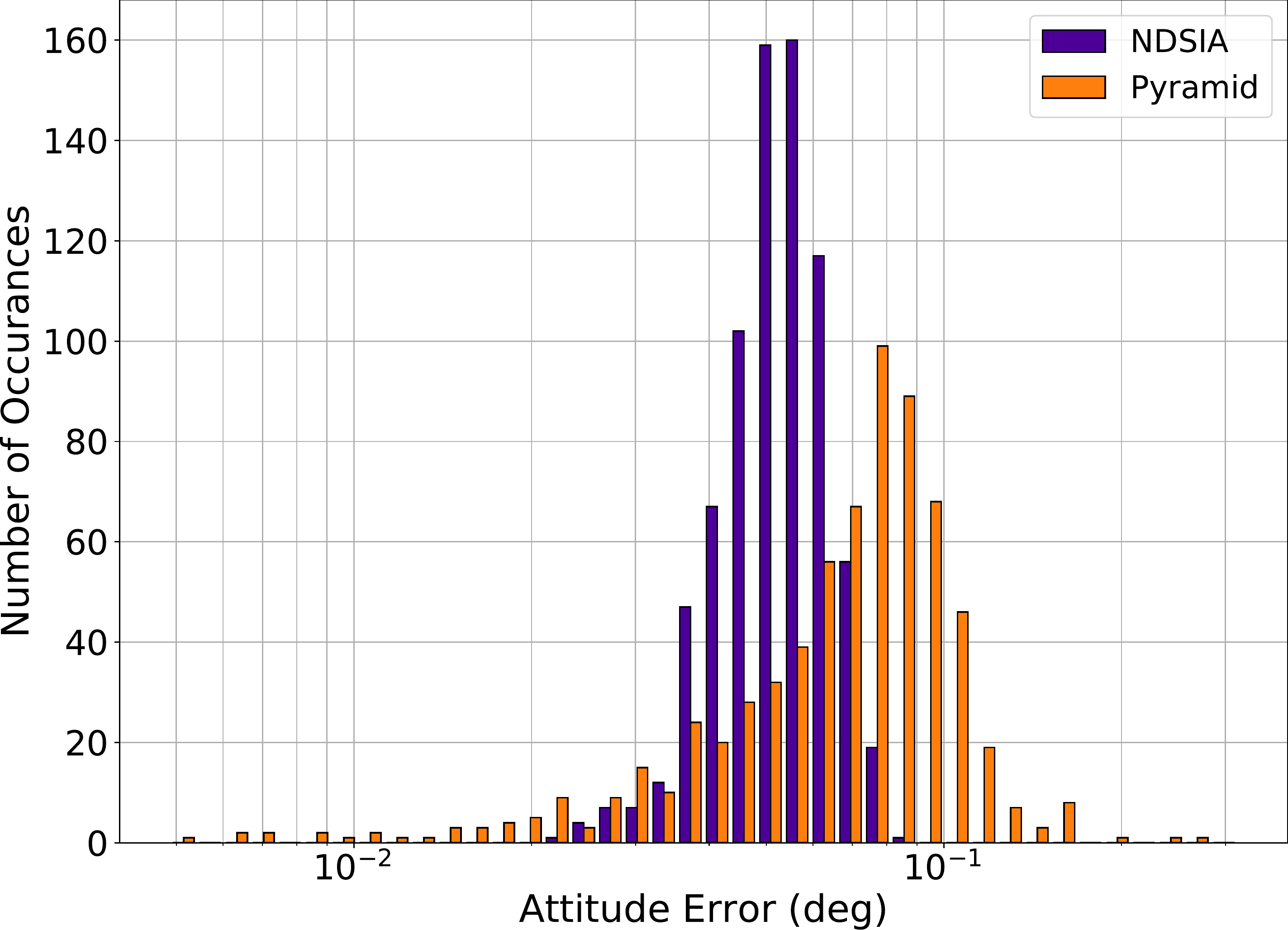}
    \caption{Test 8, Small Focal Length and OA Offset Perturbation and Larger Centroiding Error, Attitude Estimation Error for Scenes with a Successful~Star-ID.}
    \label{fig:test8}
\end{figure}

Figure~\ref{fig:test8} shows the attitude error histograms for Pyramid and the NDSIA for Test 8. This figure shows that on average the NDSIA is more accurate than Pyramid. In~Test 8, the~NDSIA was able to complete the Star-ID process more times than Pyramid. Comparing the results of this test with Test~6 reveals that the performance of the NDSIA is similar between the two tests, whereas Pyramid's performance improves noticeably in Test~8. 
\unskip

\section{Future~Work}
Although an outline for implementing the NDSIA and Pyramid in tandem is given in this article, future research should work out the finer details of said implementation. Furthermore, future research should investigate how the myriad star kernel generators, shifting necklace~\cite{triads} and greedy scene elimination~\cite{Mueller} to name a few, impact the performance of the NDSIA. In~addition, the~NDSIA considers the magnitudes of the dihedral angles that make up the spherical triangle, but~not their orientation. Calculating the orientation of the triangles as is done Reference~\cite{ND3} would add another differentiating parameter, and~thus another dimension, to~the non-dimensional database that could be used to further the confidence of the Star-ID. Moreover, increasing the dimensionality of the non-dimensional database by using three dihedral angles rather than one or two planar angles allowed the NDSIA's non-dimensional database to have a larger number of elements compared to previous algorithms. Thus, adding a fourth dimension would allow for an even larger non-dimensional database. While adding another dimension increases the difficulty of the orthogonal range search, the~$n$-dimensional $k$-vector has been shown to handle four-dimensional range searches with ease~\cite{NDKV}.

\section{Conclusions}

This work introduces the Non-Dimensional Star-Identification Algorithm (NDSIA), a~real-time star identification algorithm for star trackers on board spacecraft. The~NDSIA is devised to complement other nominal lost-in-space algorithms, for~instance Pyramid. When the nominal lost-in-space algorithm fails due to perturbations of the focal length and/or the OA offset in the star tracker, the~NDSIA is still able to identify stars. Moreover, not only can the NDSIA be used to identify stars when the nominal lost-in-space algorithm fails, it can be used to re-calibrate the star~tracker.

The idea behind the NDSIA is to identify stars via a database containing information about spherical star triangles that can be generated from a set of stars in a star catalogue. In~particular, each~element from this database contains both the stars that generated the star triangle, and~values of the dihedral angles that these stars form in a unit 3D sphere. Each time that an identification is required, three candidate stars from the frame are selected and checked against the database. If~a unique match is found, then two additional stars, called reference stars, from~the frame are selected and used to generate all nine possible star triangles between each reference star and the stars in the original triangle. If~these nine star triangles also produce unique matches in the database, then the algorithm identifies this subset of five stars. The~original three stars are used to continue with the identification of the remaining candidate stars from the image. Each remaining star is identified using a process similar to the first reference star. Finally, once the process is finished, the~interstellar angles between each pair of identified stars is used as a final check to ensure the star identification was performed~correctly.

It is important to note that in order to perform the identification of a star triangle in the algorithm database, a~three-dimensional search is required, where each dimension of the search corresponds to each one of the dihedral angles included in the database. Using a three-dimensional database significantly reduces the number of star triangle combinations for a given dihedral angle tolerance at the cost of making the searching process more difficult. Nevertheless, the~increased search difficulty is solved in the NDSIA by using the \ndkv\ (NDKV) algorithm.

Through a series of eight tests in this article, the~NDSIA was demonstrated to be robust to changes in both the focal length and OA offset of a star tracker; it provided a reliable Star-ID even when these camera parameters were perturbed. In~contrast, algorithms such as Pyramid showed an alarming increase of failures or incorrect Star-IDs when subject to focal length perturbations. Under~perturbations just in the OA offset, the~tests performed in this work showed little effect on~Pyramid.

One important property of the NDSIA is that the algorithm can be used to compute the new values of the focal length and/or OA offset of the star tracker camera. Using this information, it is possible to update the star database on board the spacecraft to take into account these perturbations. That way, the~nominal star identification algorithm, which usually has a better speed performance and lower computational cost, can be used again as it is more~efficient.

The results presented in this work show that using a combination of the NDSIA and Pyramid algorithms provides a fast and robust methodology for dealing with the real-time star identification problem even under the effect of perturbations in the focal length and/or the OA offset of star trackers. Pyramid is to be used whenever the star tracker is functioning nominally, and~the NDSIA when perturbations occur. Together, the~strengths of one algorithm complement the weaknesses of the other, increasing the reliability of the system as a~whole. 
\section*{Acknowledgments}
This work was supported by a NASA Space Technology Research Fellowship, Leake [NSTRF 2019] Grant \#: 80NSSC19K1152. The authors would like to thank an unknown reviewer for his/her insightful comments and helpful~suggestions.

\appendix
\section{Interstellar and Dihedral Angles of Spherical Star~Triangles}\label{Sec:AppI}

\begin{figure}[H]
    \centering\includegraphics[width=0.45\linewidth]{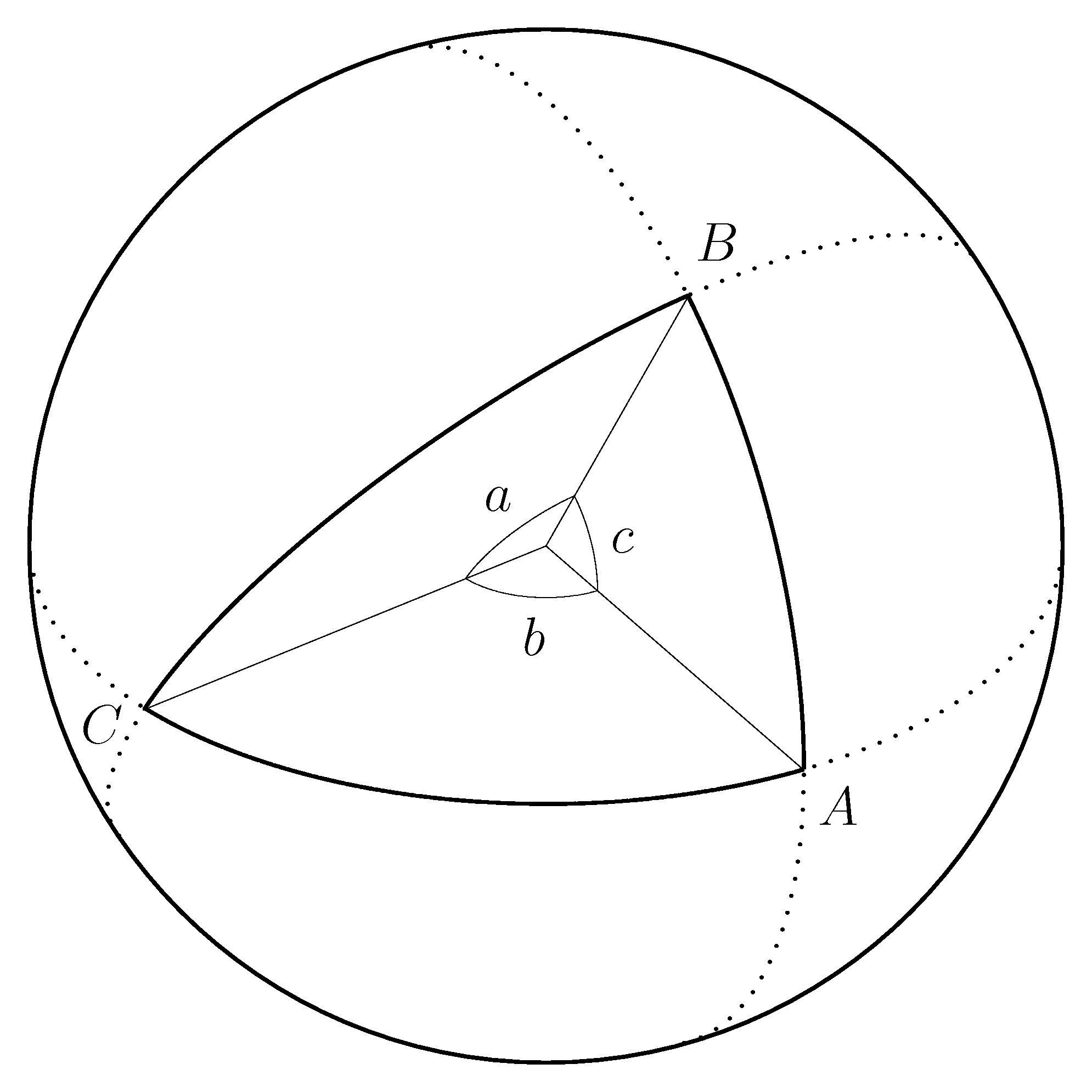}
    \caption{Spherical~Triangle.}
    \label{fig:SphTri}
\end{figure}

Figure~\ref{fig:SphTri} shows a spherical triangle. The~angles a, b, and~c are interstellar angles, the~angles between two stars. Angles $A$, $B$, and~$C$ are the dihedral angles of the spherical triangle formed by these three star~pairs.

The interstellar angles can be calculated using Equation~(\ref{Eq:InterAngle}).
\begin{equation}\label{Eq:InterAngle}
\begin{cases}
    a = \cos^{-1}\left(\hat{\B{b}}_B\T \hat{\B{b}}_C\right) \\
    b = \cos^{-1}\left(\hat{\B{b}}_A\T \hat{\B{b}}_C\right) \\
    c = \cos^{-1}\left(\hat{\B{b}}_A\T \hat{\B{b}}_B\right)
\end{cases}
\end{equation}
where $\hat{\B{b}}_A$ is the unit-vector from the origin to the star located at A, $\hat{\B{b}}_B$ is the unit-vector from the origin to the star located at $B$, and~$\hat{\B{b}}_C$ is the unit-vector from the origin to the star located at $C$.

The way the dihedral angles are calculated depends on which of the interstellar angles is the smallest. If~a is the smallest angle, then the first line of Equation~(\ref{Eq:DiHedAngCos}) is used. If~b is the smallest angle, then the second line of Equation~(\ref{Eq:DiHedAngCos}) is used. If~$c$ is the smallest angle, then the third line of Equation~(\ref{Eq:DiHedAngCos}) is used.
\begin{equation}\label{Eq:DiHedAngCos}
\begin{cases}
    A = \cos^{-1}\left(\dfrac{\cos a - \cos b \cos c}{\sin b \sin c}\right) \\
    B = \cos^{-1}\left(\dfrac{\cos b - \cos a \cos c}{\sin a \sin c}\right) \\
    C = \cos^{-1}\left(\dfrac{\cos c - \cos a \cos b}{\sin a \sin b}\right)
\end{cases}
\end{equation}

Once one of the dihedral angles is calculated using Equation~(\ref{Eq:DiHedAngCos}), the~remaining two dihedral angles are calculated using Equation~(\ref{Eq:DiHedAngSin}).
\begin{equation}\label{Eq:DiHedAngSin}
    \dfrac{\sin A}{\sin a} = \dfrac{\sin B}{\sin b} = \dfrac{\sin C}{\sin c}
\end{equation}

The attentive reader will notice that this method of calculating the dihedral angles involves an ambiguity, as~the inverse sine function will only return positive answers between $0$ and $\pi/2$, when~a dihedral angle may be larger than $\pi/2$. This ambiguity can be avoided by using a tangent rule to calculate the remaining two dihedral angles. To~test the effect of this ambiguity, all of the tests shown in this article were completed using the sine rule and the tangent rule, and~there were no differences in the percentage of Star-IDs completed nor in the percentage of Star-IDs with an attitude error less than three degrees when using the tangent rule versus the sine rule. Moreover, when using the tangent rule, the~computational time always increased compared to when using the sine rule; in one case, the~average computational time was four times larger when using the tangent rule than when using the sine rule. Thus, for~all data reported in this article, the~sine rule was always used to calculate the larger two dihedral~angles. 
\section{Pyramid Algorithm~Summary}\label{app:PyramidSummary}
Pyramid algorithm is summarized in three~steps.
\begin{enumerate}
    \item First, the~algorithm searches for a unique star
    triangle. A~unique star triangle consists of three stars whose interstellar angles, the~angle between a pair of stars, could only form that particular star triangle. In~other words, after~merging the results of three searches, one for each of the three star pairs, in~a database of interstellar angles, only one possible star triangle emerges. The~searches in the interstellar angle database are range searches where the range is based on three standard deviations of the star tracker's centroiding uncertainty. Let the three stars that make up the unique star triangle be denoted as the $i$, $j$, and~$k$ stars. If~no unique star triangle is found, then the algorithm reports that it cannot identify any stars. 
	\item Next, Pyramid searches for a reference star, $r$. A~reference star $r$ is any star in the field-of-view such that the $\{i,j,r\}$, $\{i,k,r\}$, and~$\{j,k,r\}$ star triangles are unique star triangles. If~a reference star is found, then the $i$, $j$, $k$, and~$r$ stars are considered identified. They represent the four vertices of a ``Pyramid.'' If no reference star is found, then the algorithm returns to step one to try to identify a different unique star triangle.
	\item Finally, Pyramid identifies or discards the remaining stars using the $\{i,j,k\}$ star triangle and the same technique as the reference star. For~a given star $s$, if~the $\{i,j,s\}$, $\{i,k,s\}$, and~$\{j,k,s\}$ star triangles are unique star triangles, then the star $s$ is identified. Otherwise, it is discarded. 
\end{enumerate}





\end{document}